\documentclass[prl,twocolumn,showpacs,amsmath,amssymb,superscriptaddress]{revtex4}
\usepackage[latin1]{inputenc}
\usepackage{graphicx}
\usepackage{dcolumn}
\usepackage{bm}
\begin{document}
\title{Influence of the upper critical field anisotropy on the transport properties of polycrystalline MgB$_{2}$}
\author{M.~Eisterer}
\email{eisterer@ati.ac.at}
\affiliation{Atominstitut der
Österreichischen Universitäten, 1020 Vienna, Austria}
\author{C.~Krutzler}
\affiliation{Atominstitut der Österreichischen Universitäten, 1020
Vienna, Austria}
\author{H.~W.~Weber}
\affiliation{Atominstitut der Österreichischen Universitäten, 1020
Vienna, Austria}\date{\today}
\begin{abstract}
The intrinsic properties of MgB$_2$ form the basis for all
applications of this superconductor. We wish to emphasize that the
application range of polycrystalline MgB$_2$ is limited by the
upper critical field H$_{c2}$ and its anisotropy. In wires or
tapes, the MgB$_2$ grains are randomly oriented or only slightly
textured and the anisotropy of the upper critical field leads to
different transport properties in different grains, if a magnetic
field is applied and the current transport becomes percolative.
The irreversibility line is caused by the disappearance of a
continuous superconducting current path and not by depinning as in
high temperature superconductors. Based on a percolation model, we
demonstrate how changes of  the upper critical field and its
anisotropy and how changes of flux pinning will influence the
critical currents of  a wire or a tape. These predictions are
compared to results of neutron irradiation experiments, where
these parameters were changed systematically.
\end{abstract}

 \maketitle

\section{Introduction}
The strong field dependence of the critical current densities in
MgB$_2$ represents one of the major problems for power
applications of this material, where a current density of at least
$10^{8}$ A/m$^2$ is needed. Such high current densities can be
obtained in conventional superconductors up to fields close to the
upper critical field. In high temperature superconductors
thermally activated depinning limits high current densities to
relatively small magnetic fields compared to the huge upper
critical fields of these materials. In MgB$_2$ thermal effects
should play only a minor role due to its comparatively low
Ginzburg number. Nevertheless, the critical currents become too
small for power applications at fields well below H$_{c2}$, even
at 4.2 K. Fortunately, the upper critical field of MgB$_2$ can be
rather easily enhanced by certain preparation conditions
\cite{Tak01,Flu03,Gur04}, doping
\cite{Rib03,Dou04,Wil04,Kaz05,Sum05} or irradiation
\cite{Eis02,Wan03,Put05}, and exceeds 30 T (at 0 K). Even for such
"high-H$_{c2}$" materials, the application range is limited to
around 10 T in polycrystalline samples. It was pointed out
recently \cite{Eis03} that this small application range was caused
by the upper critical field anisotropy of MgB$_2$. In the pure
material, B$_{c2}$ is about 14 T, if the field is applied parallel
to the boron planes, while it is smaller than 3 T, if the field is
applied perpendicular to the boron planes, leading to an
anisotropy of close to 5
\cite{Elt02,Ang02,Zeh02,Sol02,Lya02,Wel02}. In polycrystalline
MgB$_2$, the grains are randomly oriented and the first grains
become normal conducting at B$_{c2}^{\bot}$, which is a few times
smaller than the upper critical field of the whole sample
(B$_{c2}^{||}$), thus reducing the effective cross section and the
critical current of the conductor. In this paper the influence of
the anisotropy will be discussed quantitatively. We will
demonstrate that B$_{c2}^{\bot}$ is most important for the
application range of MgB$_2$.

\section{Distribution of the upper critical Field}

\begin{figure}
\centering \includegraphics[width = \columnwidth]{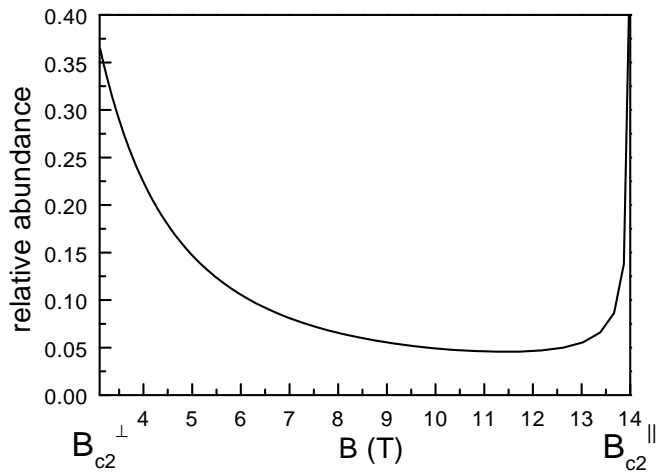}
\caption{Distribution of the upper critical field within the
grains} \label{fig1}
\end{figure}

\begin{figure}
\centering \includegraphics[width = \columnwidth]{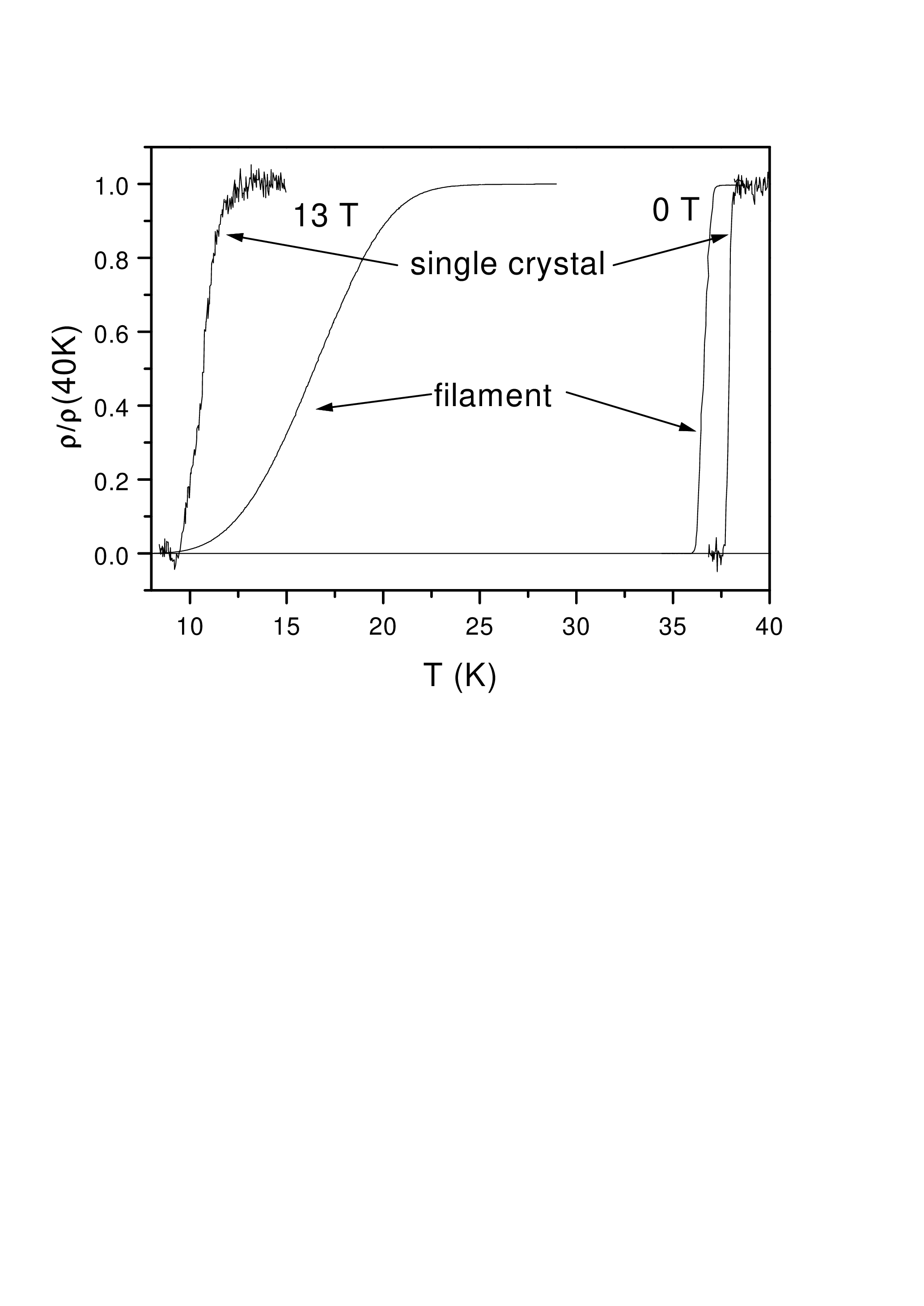}
\caption{Resistive transition of a polycrystalline sample
("filament") and of a single crystal (H$\| ab$). The applied
current density was in both specimens about 10$^{5}$ A/m$^{2}$.}
\label{fig2}
\end{figure}

The material is assumed to be perfectly homogeneous, i.e. all
grains are supposed to have identical intrinsic properties.
B$_{c2}$ of each grain depends only on its orientation to the
applied field and can be calculated within anisotropic Ginzburg
Landau theory \cite{Til65}:
\begin{equation}
B_{c2}(\theta)=\frac{B_{c2}^{||}}{\sqrt{\gamma^{2}\cos^{2}(\theta)+\sin^{2}(\theta)}}
\label{equGL}
\end{equation}
$\gamma$ denotes the anisotropy factor of the upper critical
field, i.e. $B_{c2}^{\|}/B_{c2}^{\bot}$, and $\theta$ is the angle
between the applied field and the c-axis. This relation was found
to be a good approximation for MgB$_2$ by torque measurements
\cite{Ang02,Zeh02}, although deviations by a few percent were
reported \cite{Ryd04}. For a given angular distribution of the
grains, the distribution of $B_{c2}$ can be easily calculated from
Equ.~\ref{equGL}. In untextured MgB$_2$, the grains are randomly
oriented (each crystallographic direction is equally distributed
over the elements of the steradian $d\Omega=\sin\theta d\theta
d\phi$) and their angles $\theta$ with respect to any direction
are distributed as $\sin\theta$. Assuming $B_{c2}^{\|}$ to be 14 T
and the anisotropy $\gamma$ to be 4.5, as in the pure material,
the distribution of Fig.~\ref{fig1} is obtained.
Unfortunately, more grains have a comparatively low $B_{c2}$. The
material can {\it now} be considered as inhomogeneous, since the
grains have different upper critical fields, depending on their
orientation to the applied field. This field induced inhomogeneity
can be observed by the broadening of the resistive transition in
magnetic fields (Fig.~\ref{fig2}). At zero field, where the upper
critical field plays no role, the transition of the filament of an
iron sheathed wire \cite{Gol04} is sharp and its width (0.7~K) not
much larger than the transition width (0.3~K) of a single crystal.
At higher fields, the anisotropy significantly broadens the
transition of the wire (7.5~K) compared to the single crystal
(1.8~K).

\section{Resistive Transition}

\begin{figure}
\centering \includegraphics[width = \columnwidth]{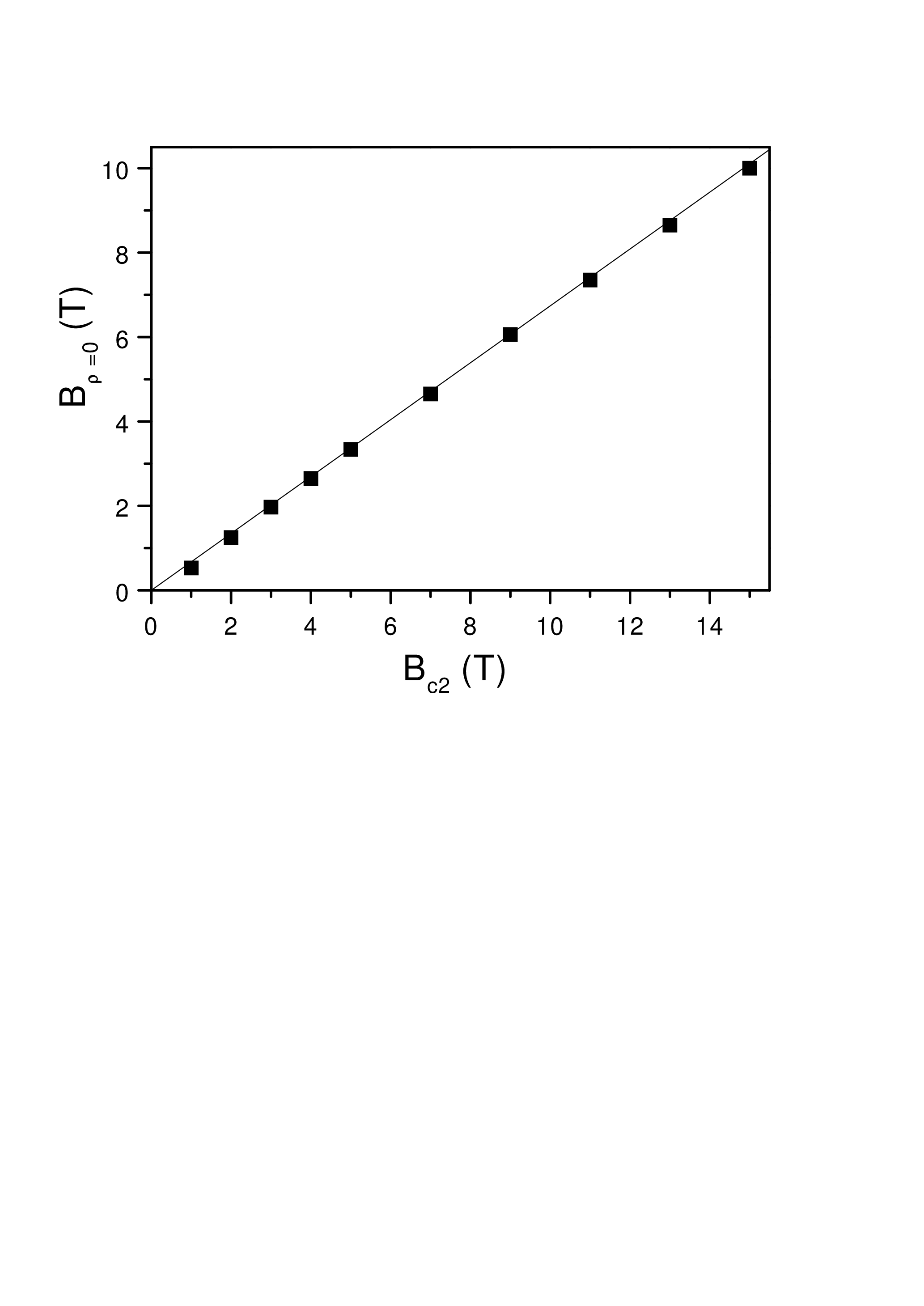}
\caption{Zero resisitivity field as a function of the upper
critical field. Symbols refer to experimental data, the line
represents the expected behavior for $\gamma$=4.5 and
$p_{c}$=0.25.} \label{fig3}
\end{figure}

If a polycrystalline sample is cooled in a fixed applied field
$B_{0}$, the resistivity starts to deviate from its normal
conducting behavior, when the first grains become superconducting.
B$_{c2}$ is defined by this onset of the transition and,
therefore, $B_{0}$ corresponds to B$_{c2}^{||}$ at the actual
temperature $T^{on}$. With decreasing temperature more and more
grains become superconducting and the resistivity decreases. A
grain becomes superconducting, if its upper critical field at the
actual temperature becomes larger than the applied field. For a
continuous superconducting current path, a certain fraction of the
grains has to be superconducting. This fraction is called
percolation threshold or critical probability $p_{c}$ and depends
on the number of connections between the grains (coordination
number). Percolation theory predicts a lower limit for $p_{c}$ of
about 0.162 for an irregular lattice \cite{Mar97} ("Finney pack")
with a coordination number of 14.3. If the coordination number is
only 6, $p_{c}$ is calculated to be about 0.31 for a three
dimensional lattice. These values correspond to the site
percolation problem, which is the appropriate model for MgB$_2$,
since the nodes (grains in the real system) are switched on or off
in site percolation problems, while the bonds between the nodes
(grain boundaries) are switched in bond percolation problems. The
fraction of superconducting grains $p$ at a field $B_{0}$, can be
calculated by integration of the distribution function:
\begin{equation}
p=\int\limits_{\theta_{min}}^{\frac{\pi}{2}}\sin{\theta}d\theta=\cos\theta_{min}
\label{equdis}
\end{equation}
In the transition region $\theta_{min}$ is defined by
$B_{c2}(\theta_{min})=B_{0}$. Inserting relation \ref{equdis} with
$p=p_{c}$ into Equ.~\ref{equGL} leads to the field of zero
resistivity
\begin{equation}
B_{\rho=0}(T)=\frac{B_{c2}^{||}(T)}{\sqrt{(\gamma^{2}-1)p_{c}^{2}+1}}
\label{equrho0}
\end{equation}
The field, where the resistivity disappears, is commonly denoted
as the irreversibility field. Since the finite resistivity in
MgB$_{2}$ above $B_{\rho=0}$ is caused by a completely different
mechanism than in high temperature superconductors, this field is
denoted as "zero resistivity field" in the following. According to
Equ.~\ref{equrho0} the zero resistivity field is proportional to
the upper critical field, as observed in experiments (Fig.
\ref{fig3}). The symbols represent experimental data obtained on
the filament of an iron sheathed wire. The onset and the offset of
the transition were defined by 90 \% and by 10 \% of the
resistivity at 40 K, respectively. The line graph was calculated
with $\gamma=4.5$ and $p_{c}=0.25$.

\begin{figure}
\centering \includegraphics[clip,width = \columnwidth]{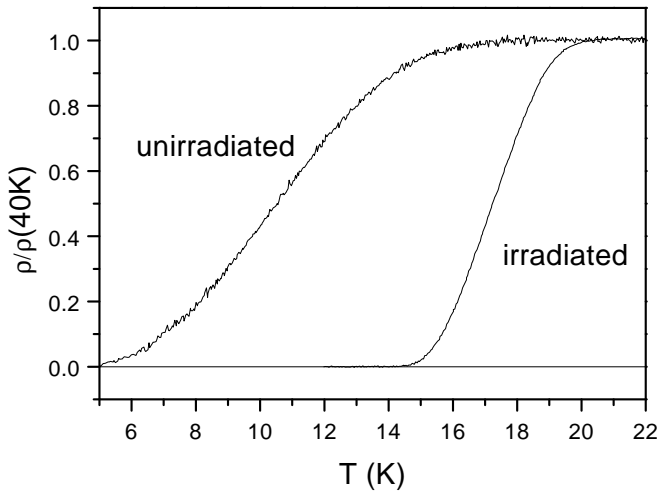}
\caption{Resistive transition at 13 T before and after neutron
irradiation.} \label{fig4}
\end{figure}

At fixed applied field the condition for the offset of the
transition becomes $B_{\rho=0}(T^{off})=B_{0}$ and - assuming only
$B_{c2}^{||}$ to be temperature dependent - the transition width
$\Delta T:=T^{on}-T^{off}$ is obtained as  \cite{Eis03}
\begin{equation}
\Delta
T=\frac{\sqrt{(\gamma^{2}-1)p_{c}^{2}+1}-1}{(-\frac{\partial
B_{c2}}{\partial T})}B_{0} \label{equdeltaT}
\end{equation}
The transition width is predicted to decrease with decreasing
anisotropy $\gamma$. This can be demonstrated by neutron
irradiation experiments. Neutron irradiation increases the upper
critical field of MgB$_{2}$ and decreases its anisotropy
\cite{Zeh04,Eis05}. The resistive transition of a sintered
MgB$_{2}$ sample is plotted in Fig. \ref{fig4} before and after
neutron irradiation. The shift of B$_{c2}$ and the predicted
decrease of the transition width can be observed.

\begin{figure}
\centering \includegraphics[clip,width = \columnwidth]{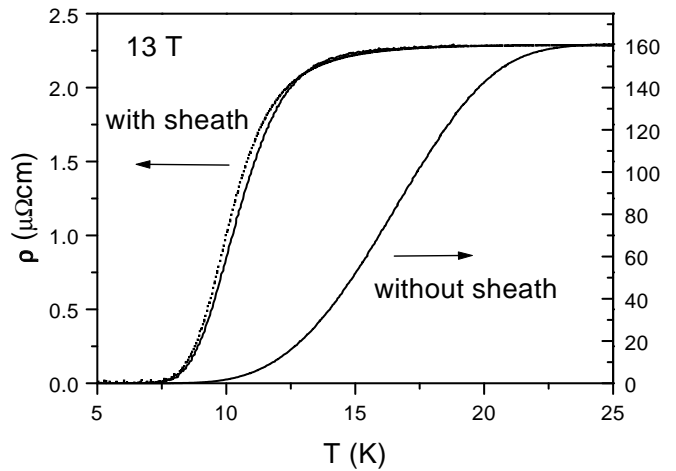}
\caption{Resistive transition of a wire before and after removing
the conducting sheath} \label{fig5}
\end{figure}

In our considerations we neglect the influence of sample
inhomogeneities and of thermal activation of the flux lines. Both
effects would additionally increase the transition width.
Inhomogeneities shift the transitions of equally oriented grains
and thermally activated depinning broadens the transition of each
individual grain, but is assumed to be zero (cf. the transition of
the single crystal in Fig. \ref{fig2}). However, in reasonably
homogeneous samples, the anisotropy is the dominant parameter,
since it predicts the correct magnitude of the transition width
(except for low magnetic fields). The transition width calculated
from Equ.~\ref{equdeltaT} represents a lower limit and is close to
typical experimental data under realistic assumptions for $p_{c}$
and for $\gamma$, not leaving much room for additional effects. On
the other hand, a conducting sheath can decrease the experimental
transition width significantly, as shown in Fig.~\ref{fig5}. The
iron sheath of a wire with a total diameter of 1.1 mm was removed
by etching, the remaining filament had a diameter of about
0.65~mm. The resistivity increased by nearly two orders of
magnitude from about 2.28~$\mu\Omega$cm to about
160~$\mu\Omega$cm. The onset of the transition (defined by the 90
\% criterion) was shifted from 12.7~K to 20~K and the transition
width increased from 3.8~K to 7.5~K. These changes are mainly due
to the removal of the parallel resistivity of the iron sheath,
since the original curve can be recalculated by assuming a
parallel resistivity of 2.32~$\mu\Omega$cm (dotted curve in
Fig.~\ref{fig5}). The small difference can be explained by the
resistivity between the iron sheath and the filament and/or by
small property changes due to the strain exerted by the iron
sheath \cite{Kov03}. For the analysis of the anisotropy, the {\it
real} B$_{c2}$ and B$_{\rho=0}$ are needed, which can only be
determined after removing the conducting sheath.

\section{Critical Currents}
\begin{figure}
\centering \includegraphics[width = \columnwidth]{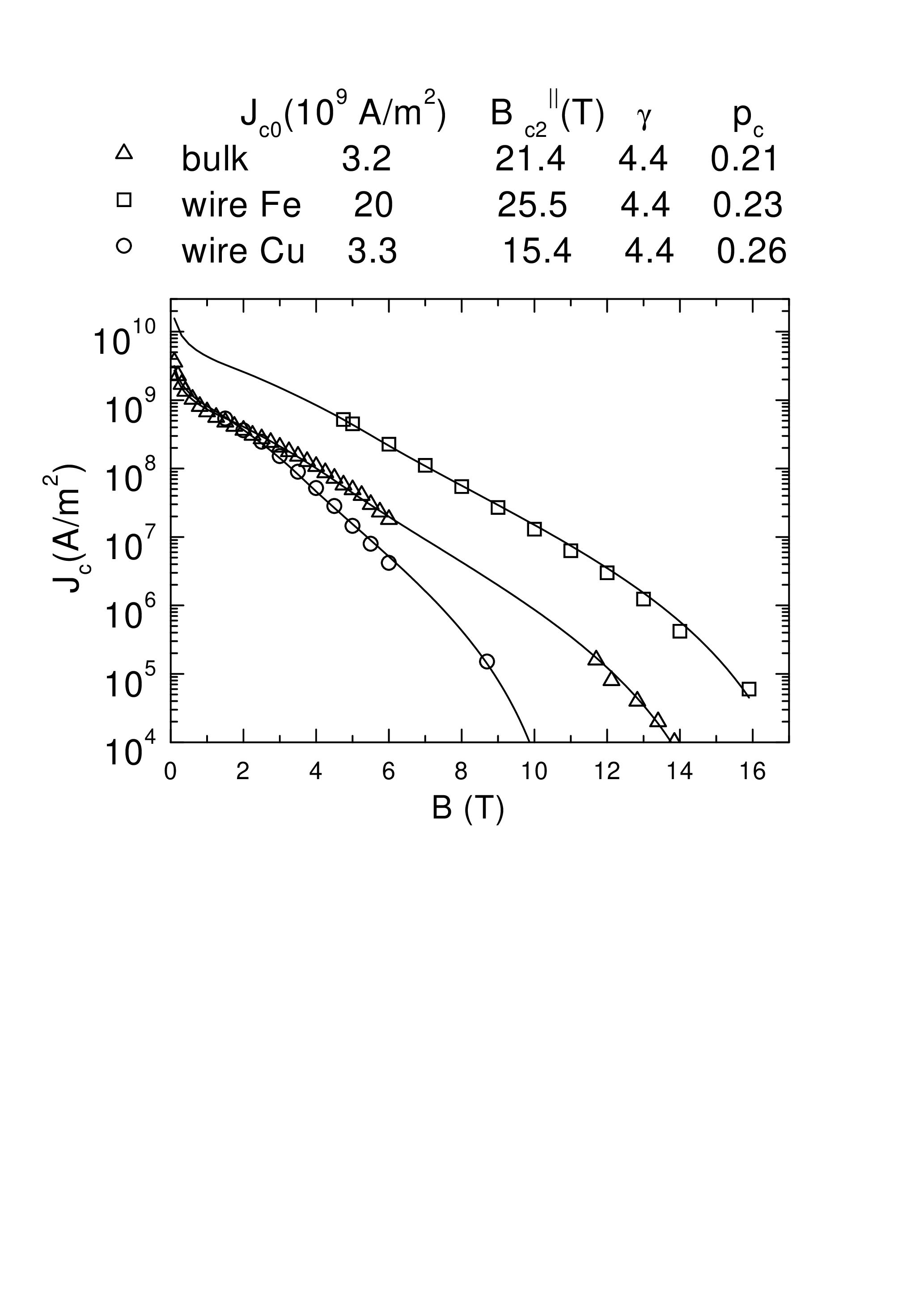}
\caption{Critical current densities of a bulk sample at 5~K and of
two different wires at 4.2~K.} \label{fig6}
\end{figure}

From the distribution of the upper critical fields (cf.
Fig.~\ref{fig1}) the distribution of the critical current
densities within the grains can be obtained, if a certain pinning
model is assumed. For grain boundary pinning, $J_{c}$ is related
to B$_{c2}$ via \cite{Dew74}
\begin{equation}
J_{c}(\theta)=J_{c0}\frac{(1-B/B_{c2}(\theta))^{2}}{\sqrt{B_{c2}(\theta)B}}
\label{equJc}
\end{equation}
for $B<B_{c2}$ and $J_{c}=0$ otherwise. The coefficient $J_{c0}$
represents the pinning strength. $J_{c}$ of a polycrystalline
sample can be calculated from the distribution of the critical
current densities \cite{Eis03} within the framework of percolation
theory. This is demonstrated in Fig.~\ref{fig6} for different
samples, i.e. a sintered bulk material, a copper sheathed in-situ
wire \cite{Glo01} and an iron sheathed in-situ wire \cite{Gol04}.
The experimental data (symbols in Fig.~\ref{fig6}) were obtained
by direct transport measurements (using a 1 $\mu$V/cm criterion)
in the case of the wires, and by ac measurements \cite{Eis00} on
the bulk sample. For the calculation of $J_{c}$ (lines in
Fig.~\ref{fig6}) four parameters are needed. The upper critical
field $B_{c2}$ of the sample (i.e. $B^{\|}_{c2}$ of the grains),
the anisotropy $\gamma$, the percolation threshold $p_{c}$ and the
pinning strength $J_{c0}$. While $B_{c2}$ was measured directly at
high temperatures and extrapolated to low temperatures, the other
parameters have to be fitted to the experimental data. The
critical current densities can be well described by the same
anisotropy for all three samples and similar values for the
percolation threshold. The pinning strength of the Cu-sheated wire
and the bulk sample is nearly identical, the smaller field
dependence of $J_{c}$ in the bulk sample is caused by its higher
$B_{c2}$. The iron sheathed wire has the largest upper critical
field and shows the strongest pinning, which leads to the highest
current densities at all fields.

\begin{figure}
\centering \includegraphics[width = \columnwidth]{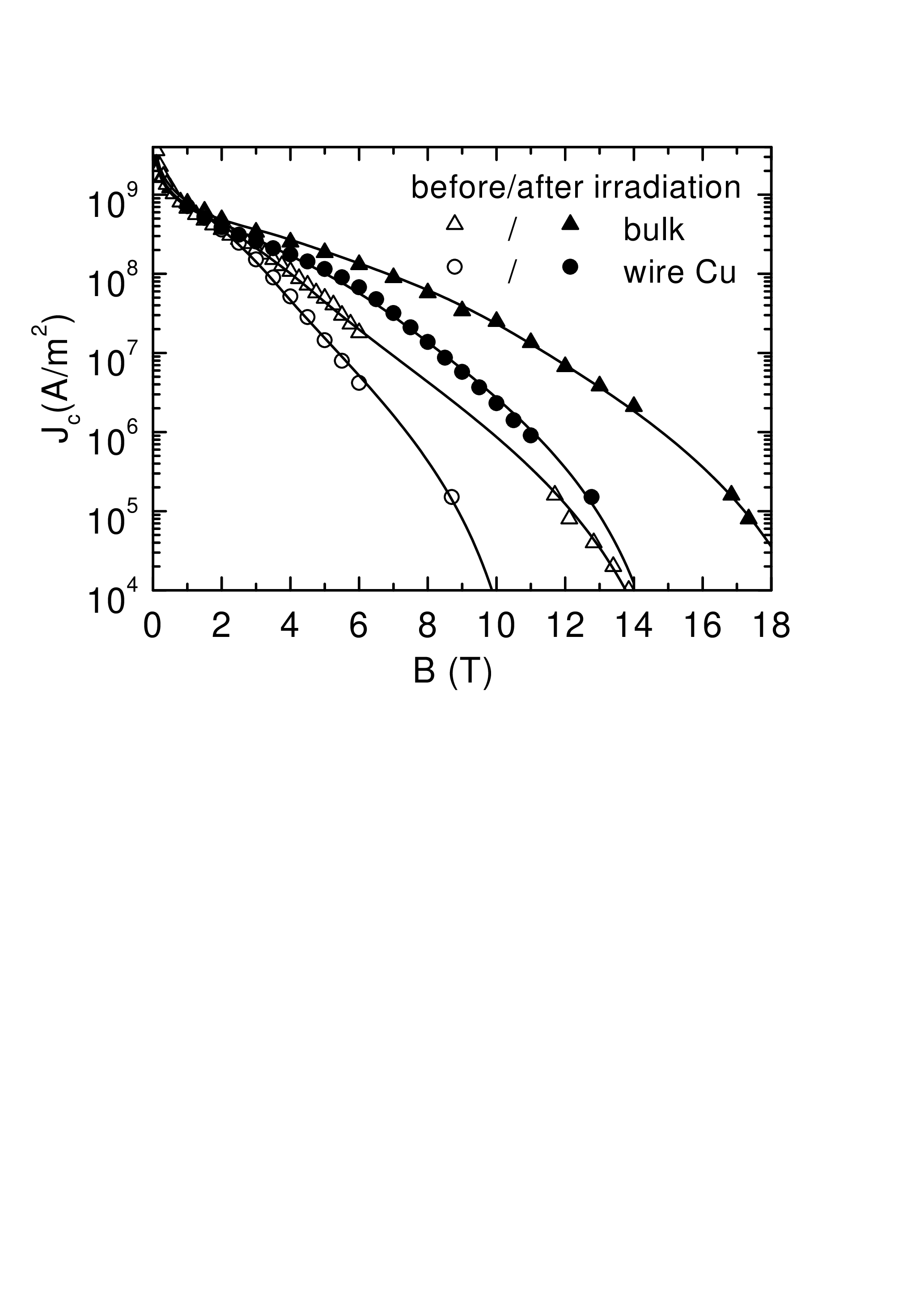}
\caption{Critical current densities before and after neutron
irradiation.} \label{fig7}
\end{figure}

The primary influcence of neutron irradiation in MgB$_{2}$ lies in
the increase of $B_{c2}$ and the decrease of anisotropy. Nearly no
changes of $J_{c}$ at low fields, but a strong enhancement of
$J_{c}$ at high magnetic fields are observed (Fig.~\ref{fig7}).
The upper critical field increases from 21.4~T and 15.4~T to 30~T
and 18.6~T, respectively, the anisotropy $\gamma$ (obtained from
fitting) decreased from about 4.4 in both samples to 2.8 and 2.6
in the bulk and in the copper sheathed wire, respectively. This
decrease of $\gamma$ is consistent with direct measurements of the
upper critical field anisotropy in single crystals
\cite{Zeh04,Eis05}. The influence on pinning is rather small. The
increase of $B_{c2}$ reduces the pinning strength \cite{Dew74}
($J_{c}\propto 1/\sqrt{B_{c2}}$), but $J_{c0}$ of the sintered
bulk sample increases by 25 \% after irradiation, which can be
attributed to the introduction of pinning centers, as observed in
single crystals \cite{Zeh04b}. These competing effects lead to an
increase of the pinning strength by only about 5 \% at low
magnetic fields. The pinning strength even decreases by about 10
\% in the wire. Therefore, the enhancement of $J_{c}$ after
neutron irradiation is caused mainly by changes in the reversible
parameters of MgB$_{2}$. Since these changes of the reversible
parameters were observed independently on single crystals, the
percolation model \cite{Eis03} correctly predicts the changes of
the critical currents in polycrystalline samples.

\section{Consequences/Strategy}
\begin{figure}
\centering \includegraphics[width = \columnwidth]{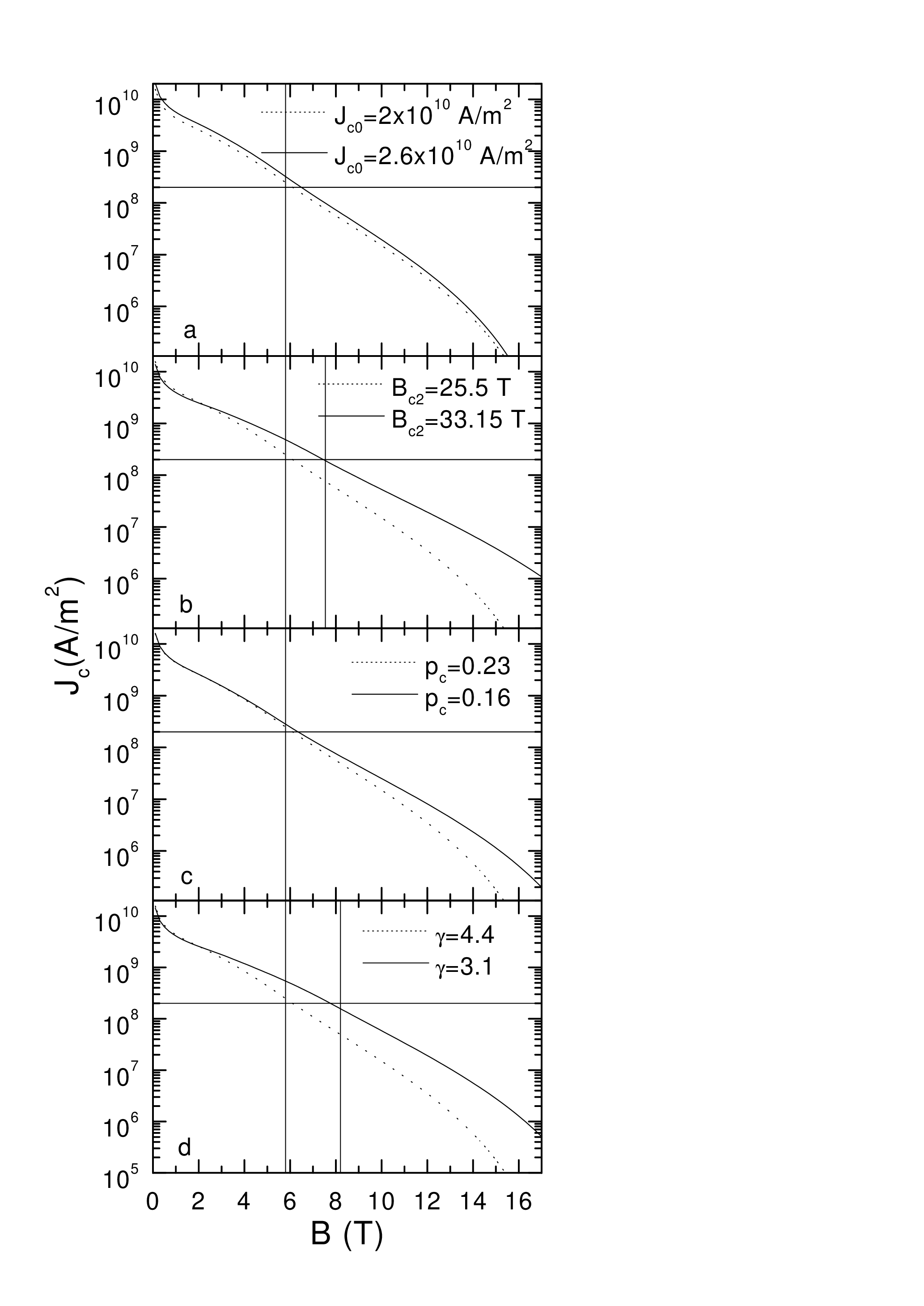}
\caption{Influence of various parameters on $J_{c}$. Only one
parameter of the original data set (dotted curve) is changed in
each panel.} \label{fig8}
\end{figure}

Based on the quantitative agreement between anisotropy-induced
percolation theory and experimental data on polycrystalline
MgB$_{2}$, the fundamental parameters of this theory can be
exploited for an assessment of their individual impact on the
current carrying capability of MgB$_{2}$. We will, therefore,
systematically investigate the influence of variations in its four
parameters. $J_{c}(B)$ calculated from the parameters of the iron
sheathed wire will be used as reference (dotted line in
Fig.~\ref{fig8}) and only one parameter will be changed by 30 \%
in each scenario. The critical current density of the wire drops
to $2\times 10^8$ A/m$^2$ (horizontal lines in Fig.~\ref{fig8}),
which is a reasonable limit for power applications and defines the
application range in the subsequent investigation, very closely to
the upper critical field $B^{\bot}_{c2}$ = 5.8 T (vertical lines
in Fig.~\ref{fig8}).

The first strategy is an improvement of pinning. This can be
achieved by the addition of pinning centers or, in the case of
grain boundary pinning, by a very fine grain structure. If the
corresponding parameter, $J_{c0}$, is increased, the critical
current densities are enhanced at all fields by the same factor
(Fig.~\ref{fig8}a). Although this is favorable at low magnetic
fields, it has little effect on the application range. The simple
shift (on a logarithmic scale) of $J_{c}(B)$ to higher values only
occurs, if the pinning mechanism remains unchanged, i.e. the
critical current densities of the individual grains can be
described by Equ.~\ref{equJc}. However, since $J_{c}$ becomes zero
at $B_{c2}$ for any pinning mechanism, the strong field dependence
and the importance of  $B^{\bot}_{c2}$ cannot be changed
qualitatively by different pinning centres.

As a second strategy, an increase of the upper critical field is
considered. It is well documented that $B_{c2}$ can be enhanced
significantly by various techniques
\cite{Tak01,Flu03,Gur04,Rib03,Dou04,Wil04,Kaz05,Sum05,Eis02,Wan03,Put05}.
Although the mechanisms involved in these changes are not yet
fully understood, impurity scattering in both bands seems to play
a major role. However, there must be a mechanism, which enhances
$B^{\|}_{c2}$ without significantly changing $\gamma$ (cf.
Fig.~\ref{fig6}). A shift of the upper critical field from 25.5~T
to 33.15~T results in a strong increase of the zero resistivity
field and in higher critical currents at high magnetic fields
(Fig.~\ref{fig8}b). Since the anisotropy is constant,
$B^{\bot}_{c2}$ increases to 7.5 T. $J_{c}$ falls below $2\times
10^8$ A/m$^2$ exactly at $B^{\bot}_{c2}$.

A smaller percolation threshold $p_{c}$ is the third possibilty to
improve the current carrying capability of polycrystalline
MgB$_{2}$. This can be achieved by a homogeneous sample with a
high density. Pores and normal conducting grains (e.g. Mg0)
increase $p_{c}$.  A small percolation threshold enhances the zero
resistivity field, but has no influence on $J_{c}$ within the
application range (Fig.~\ref{fig8}c). In principle the densitiy of
the sample also affects $J_{c0}$ due to the reduction of the
superconducting cross section, but this effect is small in samples
of good quality. For the present parameters, an increase of
$J_{c0}$ by about 15 \% is expected. Since a $p_{c}$ of 0.16
corresponds to the theoretical limit \cite{Mar97}, a significant
improvement of the transport properties cannot be expected from
the optimization of $p_{c}$.

The last strategy is a reduction of the anisotropy $\gamma$. A
reduction of $\gamma$ was reported after neutron irradiation
\cite{Zeh04,Eis05} or can be achieved by carbon doping
\cite{Mas04}, but it seems to be always accompanied by an increase
of the upper critical field $B^{\|}_{c2}$ (which further enhances
$B^{\bot}_{c2}$ and enlarges the application range). Although
experimentally difficult, the influence of such a "pure" change of
anisotropy on $J_{c}$ is demonstrated in Fig.~\ref{fig8}d.
$\gamma$ is reduced from 4.4 to 3.1 or, equivalently,
$B^{\bot}_{c2}$ is increased by 2.4~T (from 5.8~T to 8.2~T). This
also enlarges the application range by about 2~T and emphasizes
again the importance of $B^{\bot}_{c2}$ for power applications of
MgB$_{2}$ in magnetic fields.

\section{Conclusions}

The upper critical field and its anisotropy are the most important
and determining for the field of zero resistivity in
polycrystalline MgB$_{2}$, in contrast to high temperature
superconductors, where the irreversibility field is related to the
pinning properties. The maximum magnetic field  for power
applications of MgB$_{2}$ is always close to $B^{\bot}_{c2}$. It
is, therefore, essential to enhance $B^{\bot}_{c2}$ for
applications of MgB$_{2}$ at high magnetic fields.

We wish to thank Sonja Schlacher and Wilfried Goldacker from the
Forschungszentrum Karlsruhe for providing us with the iron
sheathed wire.

\end{document}